\documentstyle[12pt]{article}
\begin{document}
\begin{center}
{\large{\bf Leading Charm Production in the Interacting Gluon
Model}}
\end{center}
\smallskip
\begin{center}
{\bf F.O. Dur\~aes $^{\dag}$ } \ ,
\ {\bf F.S. Navarra $^{\ddag}$ } \ ,
\ {\bf C.A.A. Nunes $^{\S}$ } \ \\[0.2cm]
Instituto de F\'{\i}sica, Universidade de S\~{a}o Paulo\\
C.P. 20516,  01452-990 S\~{a}o Paulo, SP, Brazil \\[0.3cm]
and\\[0.3cm]
{\bf G. Wilk $^{\P}$ } \\[0.2cm]
Soltan Institute for Nuclear Studies,
Nuclear Theory Department\\
ul. Ho\.za 69, \ Warsaw, Poland

\end{center}
\date{\today}
\vspace{0.7cm}
\begin{abstract}
We discuss leading charm production in connection with energy
deposition in the central rapidity region. Special attention is
given to the correlation between production in central and
fragmentation regions. If the fraction of the
reaction energy released in the central region increases the
asymmetry in the $x_F$ distributions of charmed mesons will become
smaller. We illustrate this quantitatively with simple calculations
performed using the Interacting Gluon Model. Leading beauty production
is also considered.

\vspace{0.5cm}

\end{abstract}

\vspace{0.8cm}

\noindent
\small {$^{\dag}$ e-mail: dunga@if.usp.br} \\
\small {$^{\ddag}$ e-mail: navarra@if.usp.br} \\
\small {$^{\S}$ e-mail: cnunes@if.usp.br} \\
\small {$^{\P}$ e-mail: wilk@fuw.edu.pl} \\


\eject

Several experiments have reported \cite{DATA} a significant
difference between the $x_F$ dependence of leading and nonleading
charmed mesons. It was not possible to explain these data with the
usual perturbative QCD \cite{QCD} or with the string fragmentation
model contained in PYTHIA \cite{PYTHIA} and some alternative
mechanisms have been advanced. The most detailed data analysis,
 including predictions for asymmetry and leading particle effect at
higher energies, has been carried out with the intrinsic charm model
\cite{ICM} (ICM). In this model the essential ingredient for a good
description of asymmetries is the recombination
mechanism which binds together the intrinsic (fast) charm
quarks and the valence quarks of the projectile. Apart from this
 fast component there is a slow one, which populates predominantly
the central rapidity (low $x_F$) region and is given by perturbative
QCD. The ICM is
thus a two-component model where the central (parton fusion) and
fragmentation regions (containing intrinsic charm) components are
completely independent and added in a simple way. In
particular, there is {\it no energy conservation constraint} imposed
on both components, which would obviously result in some simple
kinematical correlations between them.

The purpose of this work is to show that such kinematical correlation
between central and non-central production is relevant for
the study of the observed asymmetries in the production of charmed
mesons and that it is also connected to another characteristic of
 high energy multiparticle production processes, namely to the
inelasticity $K$ of the reaction. It defines the fraction of the initial
energy $\sqrt{s}$ which is released and deposited in the central
region of reaction. In particular its energy ($\sqrt{s}$) dependence
will be important here. All models that adress charm production in
the central region predict that there is no asymmetry in this region.
Asymmetry comes from the fragmentation (large rapidity $y$) region.
Therefore, if $K$ increases with energy there will be less energy
available in the large $y$ region and this will result in a softening
of the leading $x_F$ distributions. Notice that this is independent of
all ingredients of the hadronization process since they are universal
and energy independent (like, for example, the fragmentation functions).
The $x_F$ distributions of the leading charmed particles will thus
eventually merge with the distribution of the centrally produced
charmed particles, which will then become broader. The asymmetry will
then not be observed! The opposite might also be true if $K$ decreases
with energy. In this case the leading system will carry
proportionally more and more energy, implying a faster leading charm
and resulting in stronger asymmetry \cite{FOOT1}.

The asymmetry problem can therefore be well formulated just in terms
of kinematical considerations. All dynamics will show itself only in
the way through which initial energy of projectiles will be
distributed in rapidity space. For example, one would naively expect
that, if perturbative QCD becomes more important at higher energies
(due to, for example, increased minijet activity), the central
production (and also energy deposition in central region) will become
dominant and the asymmetry will decrease or even disappear. This goes
along with the expected increase with energy of inelasticity deduced
from the analysis of accelerator and cosmic ray data \cite{INEL}.

In what follows we shall therefore study
 leading charm production in terms of the Interacting Gluon Model
(IGM) \cite{IGM}, which has been invented to describe the
inelasticity and its energy behaviour and recently used also to
successfuly describe many aspects of multiparticle production
(including its semi-hard minijets component, which can be important
for charm production at high energies).

Since the IGM has already been described previously in great detail
\cite{IGM} we shall provide here, for completeness, only the most
basic formulas and concentrate our attention on the specific
mechanisms of charm production and on the calculation of the
asymmetries between leading and nonleading charm mesons.
The asymmetry has been most accurately
measured in the $\pi p$ scattering, therefore we shall start
discussing this process first \cite{FOOT2}.
In Fig. 1a  we show the IGM description of the energy flow in a hadron-
hadron collision at high energies. Through the cooperative action of a
certain number of soft gluons, carrying an overall momentum fraction $x_1$
of the incoming pion, colliding with a similar bunch of gluons coming
from the target nucleon and carrying fraction $x_2$ of its momentum,
an object called central fireball (CF) is formed. It will decay later
on producing observed secondaries. In the IGM \cite{IGM} the probability
for this to happen is given by the function $\chi(x_1,x_2)$.
The pion remnants leaving the central region (i.e., their valence
quarks plus some gluons which did not interact) carrying momentum
$x_L$ are called in the IGM the {\it leading jet} (LJ) and, being
themselves excited objects, may also produce particles (including
$D$ mesons).

\eject

{}From the basic function $\chi(x_1,x_2)$ we can compute
the Feynman momentum distributions of the CF, $\chi(x_{CF})$, where
$x_{CF}=x_1-x_2$, and of the LJ, $f_{LJ}(x_L)$, by a simple change of
variables:
\begin{eqnarray}
\chi(x_{CF}) & = & \int^1_0 dx_1 \int^1_0 dx_2 \, \delta\left(
           x_{CF}- x_1 + x_2 \right) \, \chi(x_1,x_2) \, \theta\left(
           x_1 x_2 s - 4 m_D^2 \right)  \label{CCF}
          \\[0.3cm]
f_{LJ}(x_L) & = & \int^1_0 dx_1 \int^1_0 dx_2 \, \delta\left(
         1 - x_1 - x_L \right) \, \chi(x_1,x_2) \, \theta\left(
         x_1 x_2 s - m_0^2 \right)
     \label{FCL}
\end{eqnarray}
where $m_D$ (1.8 GeV) and $m_0$ are the masses of the $D$ meson and
of the lightest state produced in such collisions \cite{IGM}. In the
above equations we clearly see the connection between central and
fragmentation production. The momentum distributions of the systems
which will later give origin to charmed particles are derived from
the same quantity $\chi(x_1,x_2)$. Moreover, $\chi(x_{CF})$ and
$ f_{LJ}(x_L)$ carry all the energy ($\sqrt{s}$) dependence of the process,
which is both explicit, in the theta functions, and implicit,
since $\chi(x_1,x_2)$ depends on $\sqrt{s} $ . In Fig.  1b we show
central $ D \overline D $ meson production where $D$($\overline D$)
is any $D$ meson carrying a $c$($\overline c$) quark.
Notice that, in the spirit of IGM, the central production
ignores the valence quarks of target and projectile (defined here as
those which carry the essential quantum numbers of the colliding
pions and protons) which, in the first approximation, just ``fly
through''. Because of this, the centrally produced $D$'s will not
show any leading particle effect.
There are two distinct ways to produce $D$ mesons out of
LJ's: fragmentation and recombination. It is assumed here that,
whenever energy allows, we shall have also $\bar{c}c$ pairs in the LJ
(produced, for example, from the remnant gluons present there). These
charmed quarks may undergo fragmentation into $D$ mesons, as shown in
Fig. 1c, but may also as well recombine with the valence quarks as
depicted in Fig. 1d. It turns out that only this last process will
produce asymmetry.
In the case of pion-nucleon scattering, the measured leading charmed
mesons are $D^-$ and the nonleading are $D^+$.

\eject

We shall now write the
Feynman $x_F$ single inclusive
distribution of $D^-$ mesons produced by the CF, by the
fragmentation in the LJ (F) and by the recombination there (R):
\begin{eqnarray}
\frac{d\sigma^{CF}}{dx_{D^-}}\, &=&\, \int^1_{x_{D^-}}\! dx_{CF}\, \,
\chi(x_{CF})\,
\int^{x_{CF}}_{x_{D^-}}\! dx_{\bar{c}}\, \,
g(x_{\bar{c}})\, D\left( \frac{x_{{D}^-}}{x_{\bar{c}}}\right) \,
\quad , \label{eq:CF} \\[0.3cm]
\frac{d\sigma^{F}}{dx_{D^-}}\, &=&\, \int^1_{x_{D^-}}\! dx_{L}\, \,
f_{LJ}(x_L)\,
\int^{x_L}_{x_{D^-}}\! dx_{\bar{c}}\, \,
     g(x_{\bar{c}})\, D\left( \frac{x_{{D}^-}}{x_{\bar{c}}}\right) \,
\quad, \label{eq:F} \\[0.3cm]
\frac{d\sigma^R}{dx_{D^-}}\, &=&\, \int^1_{x_{D^-}}\! dx_L\, \, f_{LJ}(x_L)\,
                               \, \int\! dx_c\, \int\! dx_{\bar{c}}
                               \, \int\! dx_{\bar{u}}
                               \, \int\! dx_d\, \, g(x_c)\, g(x_{\bar{c}})\,
             f(x_{\bar{u}})\, f(x_d) \nonumber \\[0.3cm]
        && \cdot\, \,
            \delta(x_{D^-} - x_{\bar{c}} - x_d)\,
             \delta(x_L - x_{\bar{c}} - x_c - x_d - x_{\bar{u}}) \quad ,
\label{eq:R}
\end{eqnarray}
where $f(x)$ and $g(x)$ are distribution functions of valence and
charm quarks respectively and the
$D(z)$'s are charm quark fragmentation functions \cite{ICM}.
The $D^+$ momentum
distributions are given by (\ref{eq:CF}) and (\ref{eq:F}) with the
replacements: $D^- \rightarrow D^+$, $ \overline c \rightarrow c$.
 These nonleading mesons will not be produced by recombination (eq.
(\ref{eq:R})). The functions $f(x)$ and $g(x)$ are essentially
unknown since they are momentum distributions of partons inside the
CF and inside the LJ after the collision. Following our assumption
that the valence quarks interact weakly we shall approximate $f(x)$
by the initial state valence quark distributions and take them from
reference \cite{DO}. As for the charm quark distribution, $g(x)$,
the situation is less clear. The $c - \overline c $ pairs do not come
directly from the sea : in the CF they are produced and in the LJ they
may be excited. It is therefore reasonable to think that the charm
quarks will be somewhat faster than ordinary sea quarks. Accordingly
we shall use for $g(x)$ the ansatz proposed by Barger and collaborators
\cite{HAL}:
\begin{eqnarray}
g(x)\,=\,\left(\,\frac{1-x}{x}\,\right)^{1/2}
     \label{GGX}
\end{eqnarray}
which is less singular than $1/x$ but still much softer than an intrinsic
charm distribution which behaves typically like $x(1-x)$. The
fragmentation functions have the Peterson form \cite{PET}:
\begin{eqnarray}
D_{c \rightarrow D}(z)=\frac{N}{z[1-1/z-\varepsilon/(1-z)]^2}
     \label{DDP}
\end{eqnarray}
where $\varepsilon \simeq \frac{\langle m_q^2+p^2_{qT} \rangle}
{\langle m_Q^2+p^2_{QT}\rangle}$ and $m_q$, $p_{qT}$, $m_Q$, $p_{QT}$
are mass and transverse momentum of the light and of the heavy quark
respectively and $N$ is a normalization constant. In the present case
$\varepsilon \simeq 0.06$.
In Fig. 2a we show the (unnormalized) contributions
coming from the three processes above (eqs. (\ref{eq:CF}),
(\ref{eq:F}) and (\ref{eq:R})). As expected, central production
(solid line) leads to the softest $D$ meson $x_F$ distribution,
 recombination in the leading jet (dotted line) leads to the
hardest final distribution and leading jet fragmentation
(dashed line) lies in between. This is so because $\chi(x_{CF})$
is softer than $f_{LJ}(x_L)$ and because recombination adds momenta
whereas fragmentation causes always some deceleration. Note that,
 although flat, the dashed and dotted curves have a pronounced
maximum at very low $x_D$. This is a direct consequence of the
behaviour of $g(x)$. If instead of the form (6) we
use an intrinsic charm distribution we will obtain a strong
suppression at low $x$ and a maximum around $x_D = 0.4 - 0.6 $ .
A final comment on this figure is that our distribution of centrally
produced $D$'s (solid line) is broader than the one obtained from
perturbative QCD. This is so because the cooperative mechanism
adds together soft gluons, increasing the energy released in the
central region, favouring higher values of $x_1$ and $x_2$ (in Fig. 1a)
and allowing for fluctuations with higher $x_{CF}$. Considering all
that was said above we can conclude that the IGM (like the ICM) is a
two component model in which the components are not very much different
in shape from each other (in sharp contrast to what happens with the
components of the ICM) and have some overlap. Because of this we expect
to find smaller asymmetries than those found in ref. \cite{ICM}, but
this depends, of course, on how one mixes the different components.
In what follows we write the differential cross section as the sum of a
central fireball (CF) and leading jet (LJ) component and the last one
as the sum of a fragmentation (F) and a recombination (R) component,
using a similar notation as in ref. \cite{ICM}:
\begin{eqnarray}
\frac{1}{\sigma}\,\frac{d\sigma}{dx_{D^-}}\, &=&\,\left(1-\eta\right)\,
\frac{1}{\sigma^{CF}}\,\frac{d\sigma^{CF}}{dx_{D^-}}\,+ \,\eta\,
\frac{1}{\sigma^{LJ}}\,\frac{d\sigma^{LJ}}{dx_{D^-}}
\quad , \label{eq:SDF} \\[0.3cm]
\frac{1}{\sigma^{LJ}}\,\frac{d\sigma^{LJ}}{dx_{D^-}}\, &=&\,
\left(1-\xi\right)\,
\frac{1}{\sigma^{F}}\,\frac{d\sigma^{F}}{dx_{D^-}}\,+ \,\xi\,
\frac{1}{\sigma^{R}}\,\frac{d\sigma^{R}}{dx_{D^-}}
\end{eqnarray}
where the mixture parameters are $\xi$ ($0\leq\xi\leq1$) and
\begin{eqnarray}
\eta\,=\,\frac{\sigma^{LJ}}{\sigma^{CF}+\sigma^{LJ}}
\end{eqnarray}
In the case of the $D^+$ distribution, the expressions above are the
same but  $\xi = 0$.

 The ICM parameter  $\eta$ has been chosen $0.2$ because of an
analogy between $\sigma_{ic}$ (our $\sigma^{LJ}$) and the diffractive
charm cross section. On the other hand, in the Valon Model \cite{VM}
the same data are adressed without any central component. This would
correspond to taking $\eta\,=\,1$.
Here, because of the kinematical mixing between CF and LJ the value
of $\eta$  is essentially free. In what follows we will choose it to be
$\eta\,=\,0.7$.
 Note also that, in our case, $\xi
=0$ corresponds to no asymmetry.
Since existing data on open charm production \cite{DATA}
apparently  do not show  nuclear effects \cite{KR}, we use here (as all
other models which address these data) the IGM for hadron-nucleon
collisions \cite{FOOT3}.\\

In Fig. 3 we compare our calculations
 with WA82 data. Fig. 3a and
3b show the $x_F$ spectrum of $D^+$ and $D^-$, respectively, and Fig.
3c shows the asymmetry which is given by:
\begin{eqnarray}
A(x_F)\, &=&\,\frac
{\frac{d\sigma}{dx_{D^-}}\,- \,\frac{d\sigma}{dx_{D^+}}}
{\frac{d\sigma}{dx_{D^-}}\,+ \,\frac{d\sigma}{dx_{D^+}}}
\end{eqnarray}

In Fig. 3b and 3c solid, dashed and dotted lines correspond to
$\xi= 0.8 ,0.5 $ and $0.2$ respectively. Data points are from the
WA82, E769 and E791 \cite{E791} collaborations.
As it can be seen, a satisfactory description of data can be
obtained with the IGM. The best description can be obtained with a
large ammount of recombination ($\xi=0.8$). This is ultimately due
to our choice of $g(x)$. We have checked that the choice of an
ordinary sea distribution for the charm quarks in the CF and LJ
requires $\eta=1.0$ and $\xi=1.0$ for a reasonable fit. The choice of
an intrinsic charm distribution allows for small values of $\eta$
and $\xi$. The conclusion seems to be that although data do not
rule out usual sea distributions as an input, good fits with more
reasonable values of the parameters can be obtained using harder
charm quark distributions like (6) or the one used in ref. 4.

We consider now the energy dependence of the asymmetry.
All details concerning the particularities of charm production
are energy independent.
In equations
(8) and (9) $\eta$, $\xi$ and the differential distributions, i.e.,
respectively normalization and shape of the curves,
can depend on $\sqrt{s}$. For simplicity we shall assume that
 $ \xi $ does
not change with the energy. The distributions $\frac{d\sigma}{dx_{D^-}}$
will depend on $\sqrt{s}$ through $\chi(x_{CF})$ and $f_{LJ}(x_L)$. The
behaviour of these last functions with $\sqrt{s}$ is shown in Fig. 4a
and 4b respectively. We observe a very modest broadening of
$\chi(x_{CF})$ implying a small increase of $\langle x_{CF} \rangle $
and a more pronounced softening of $f_{LJ}(x_L)$ with the corresponding
reduction of $\langle x_L \rangle $. As for $\eta$, an extensive
analysis \cite{IGM} of charged particle
production up to Tevatron energies has shown that it decreases
by a factor 3   when we go from ISR to Tevatron energies.
Assuming a similar reduction for the case of charmed particle
production $\eta$ will change from 0.7 to 0.25.
Considering what was said above we evaluate
again all the expressions (1)-(9) at $\sqrt{s} =1800$ GeV. The
resulting asymmetry is shown in Fig. 4c with a dashed line. For
 comparison we show in the same figure (with a solid line) the
asymmetry at $\sqrt{s}=26$ GeV calculated with the same parameters. It
decreases $20\%$ in the region $x_F \geq 0.5$. Although this is not
a very impressive change it illustrates the trend. Moreover, we know that
the asymmetry goes asymptotically to zero since $\eta \rightarrow 0 $.
We emphasize that this is so because the IGM
(in its version \cite{IGM}) predicts that at higher energies, because of
the action of minijets, the energy deposition in the central region
will increase implying two effects:
 a growth of the central multiplicity ( implying
 thus an increase of $1-\eta$ ) and
  a softening of the leading jet momentum
 distribution.
We can therefore conclude that,
irrespective of details of charm production, these both effects
combined will reduce the asymmetry. It is interesting to mention that
the data collected in Fig. 3c come from three different
collaborations E769, WA82 and E791 with beam energies of 250, 340 and
500 GeV respectively. In the CMS this corresponds to a variation of
$\sqrt{s} = 23 $ GeV to $\sqrt{s} = 33$ GeV. This energy change is
small, the error bars are large and therefore no change in the asymmetry
is visible yet. At higher energies there is a chance to experimentally
verify this behaviour at RHIC or LHC.

As a straightforward extension of our analysis we calculate now the
asymmetry in $B$ meson production. This is done by simply replacing
$m_{D}$ by $m_{B}= 4.75$ GeV in (1) and $\varepsilon = 0.06$ by
$\varepsilon= 0.006$ in (7). In principle we should also change
$g(x)$ but in a first estimate we keep the ansatz (6). If we would
use an intrinsic distribution for $g(x)$ it would be very weakly
dependent on the heavy quark mass \cite{ICM}. In Fig. 5 we show
the $\chi(x_{CF})$ distribution for charm
(dashed line) and for beauty production (solid line) with the
proper change  in eq. (1). The energy is
$\sqrt{s}=26$ GeV. The effect of increasing the production threshold
($m_D \rightarrow m_B$ in the theta function in eq.(1)) is to  select
events with a more massive CF and with larger lower limits for
$x_1$ and $x_2$, suppressing thus larger values of $x_{CF} = x_1-x_2$
with respect to charm production (in the limit of total energy
deposition, i.e., $x_1=x_2=1$, the CF would be at rest). This effect
is however very small. This is expected and seen in Fig. 5.
In Fig. 6a (6b) we show the $x_F$ distributions of nonleading
(leading)$D$ and $B$ mesons. The energy is the same as in Fig. 5 and
the parameters
are the same as before ($\eta=0.7$ and $\xi=0.8$). Nonleading
spectra are calculated with eqs. (3) and (4). The Peterson
fragmentation functions, appearing in those equations, are very
sensitive to the value of $\varepsilon$. In the case of beauty,
the strong reduction of $\varepsilon$ makes the fragmentation
function strongly peaked at very large values of z. The emerging
$B$'s will therefore be much less decelerated than the $D$'s. This
effect compensates the previous one and the final nonleading $B$
distribution is harder than the nonleading $D$ one. The leading
distributions include recombination, given by eq. (5), which
is not affected by the change in the heavy quark mass. Because
of this, the spectra in Fig. 6b exhibit the same qualitative
behaviour ($B$'s faster than $D$'s) seen in Fig. 6a but the
difference between $B$'s and $D$'s is smaller. The asymmetries
of $B^-$/$B^+$ and $D^-$/$D^+$ are shown in Fig. 7  with
solid and dashed lines respectively  for $\sqrt{s}=26$ and $1800$
GeV.  The asymmetry in beauty is about $50\%$ weaker than in charm
at $x_F=0.8$ and both show a similar decrease with energy.

In conclusion: the IGM describes the energy flow in high energy
hadron collisions. It takes properly into account the correlation
between energy deposition in the central region and the leading
particle momentum distribution. It accounts for
 charmed meson production in a natural
and satisfactory way and makes the prediction that at higher energies
the increase of inelasticity $K$ (see \cite{IGM}) will lead to the
decrease of the asymmetry in heavy quark production. It also predicts
a weaker asymmetry for beauty. We believe that this point should also
be adressed by other models which deal with asymmetry in heavy flavour
production \cite{KOD}.

\vspace{1cm}

\noindent {\bf ACKNOWLEDGEMENTS}

This work was supported partially by FAPESP, by CNPq and by the Polish State
Committee for Scientific Research Grant. One of us (F.S.N.) is deeply
indebted to his Polish colleagues for the warm hospitality extended to
him during his stay in Warsaw. It is a pleasure to thank
R. Vogt for very instructive discussions.

\eject

\vspace{1cm}

\eject
\noindent
{\bf Figure Captions}\\

\begin{itemize}
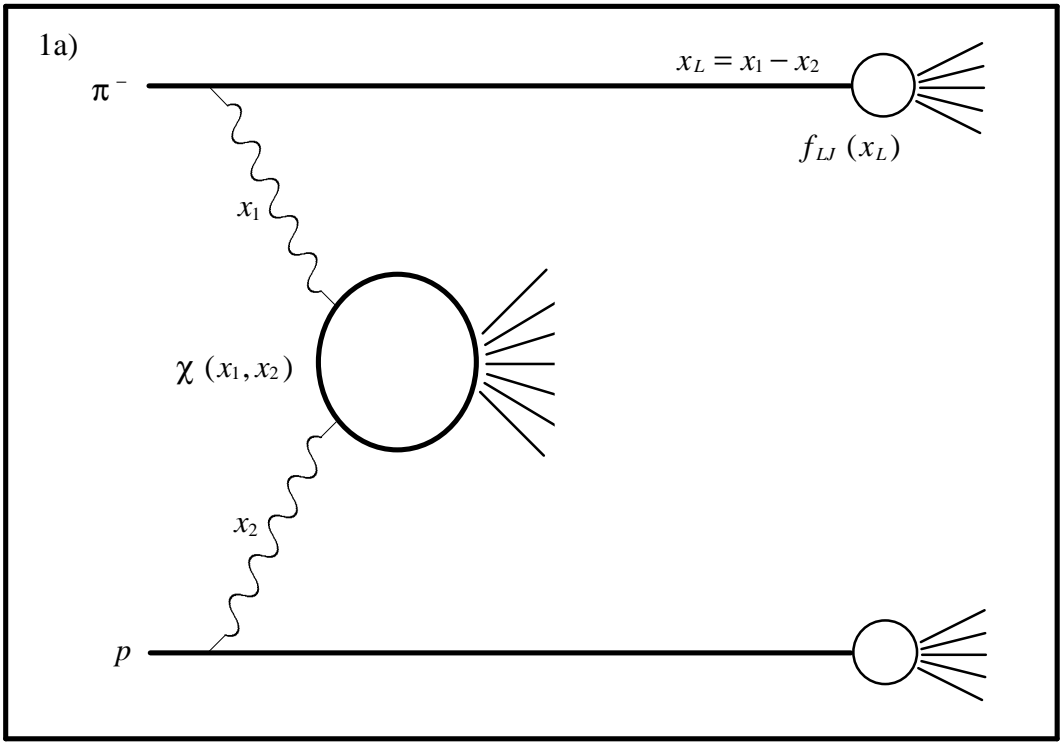
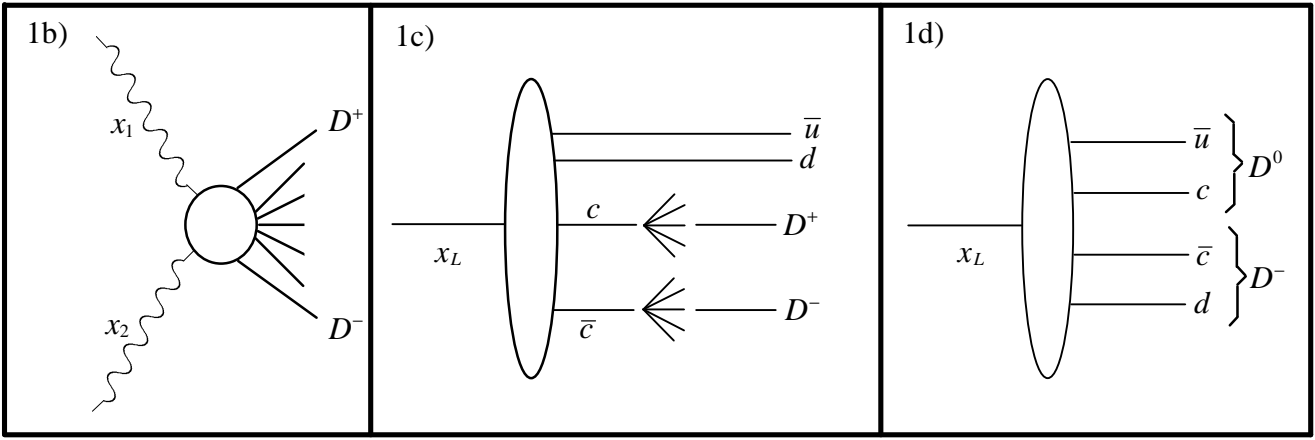
\item[{\bf Fig. 1}] $(a)$ Illustration of a pion-nucleon collision:
fractions $x_1$ and $x_2$ of the incoming hadrons momenta form a
central fireball (CF) with probability $\chi(x_1,x_2)$. The fraction
$1-x_1 = x_L$ is carried by the leading jet (LJ). The leading jet
momentum spectrum is $f_{LJ}(x_L)$; $(b)$
Nonleading $D$ meson production by central fireball
fragmentation; $(c)$ Nonleading $D$ meson production by leading jet
fragmentation; $(d)$ Leading $D$ meson production by leading jet
recombination.
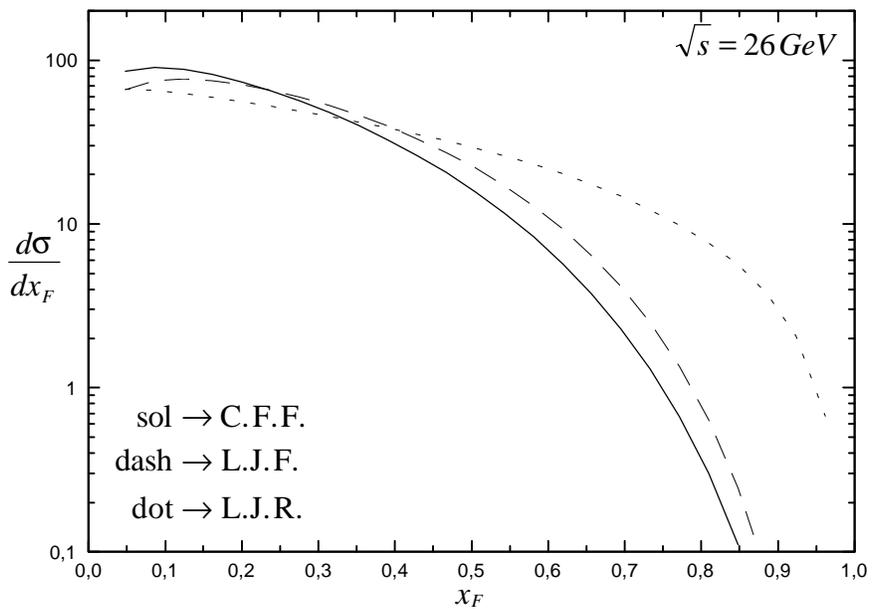
\item[{\bf Fig. 2}] $x_F$ distributions of $D$ mesons calculated
 with the IGM.
The solid line shows the contribution of the central fireball
fragmentation. The dashed line shows the contribution of the leading
jet fragmentation and the dotted line shows the contribution of the
leading jet recombination.
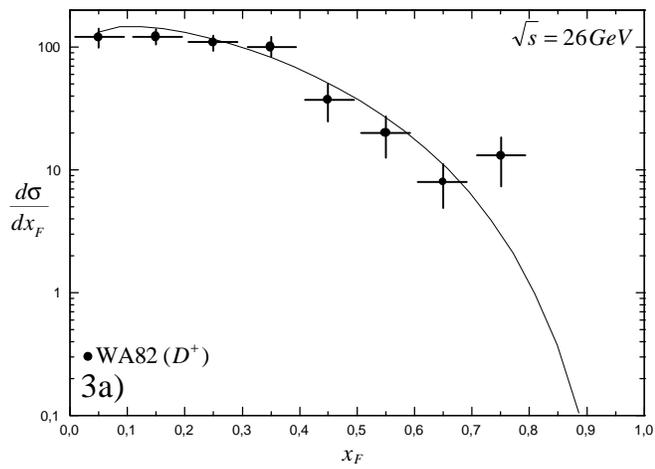
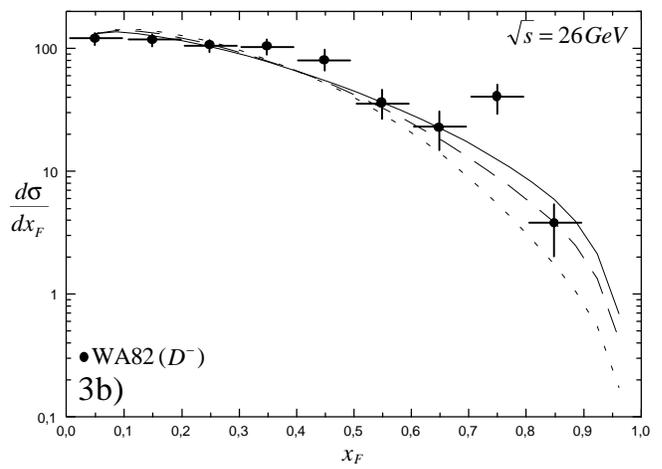
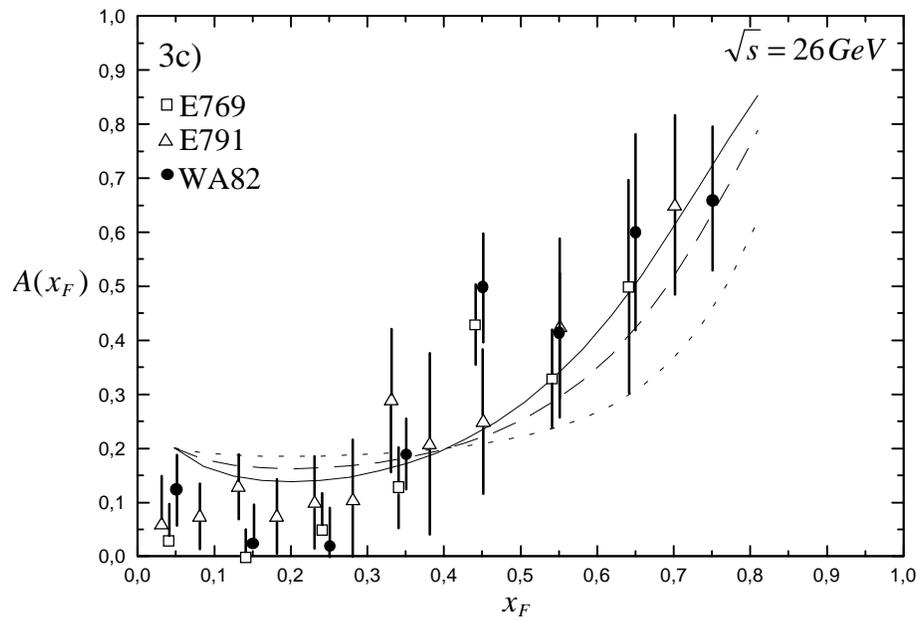
\item[{\bf Fig. 3}] $(a)$  $x_F$ distribution of $D^+$ mesons
calculated with our model and compared with WA82 data; $(b)$ $x_F$
distribution of $D^-$ mesons calculated with our model and conpared
with WA82 data. Solid line corresponds to $\xi=0.8$ in eq. (9)
while dashed  and dotted lines correspond to $\xi=0.5$ and $\xi=0.2$
respectively; $(c)$ the asymmetry calculated with the
IGM and compared with WA82 (solid circles), with E769 (open squares)
and E791 (open triangles) data. Solid, dashed and dotted lines
correspond to the same choices of $\xi$ of (3b).
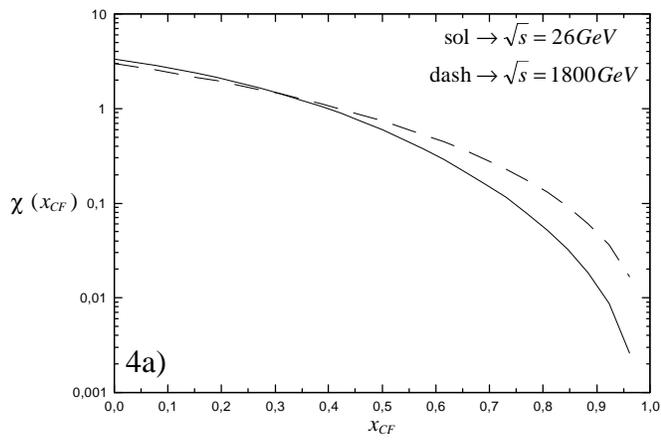
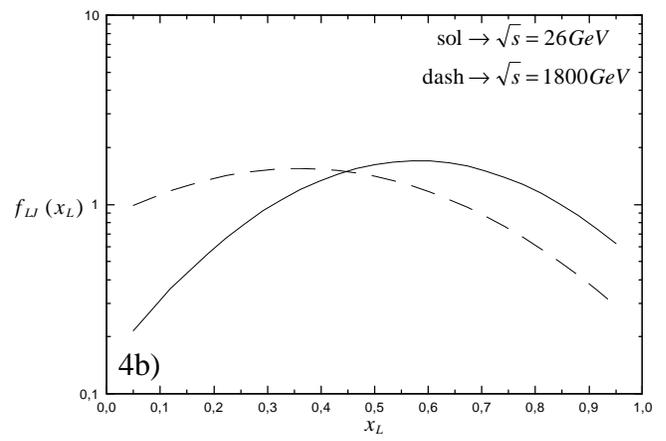
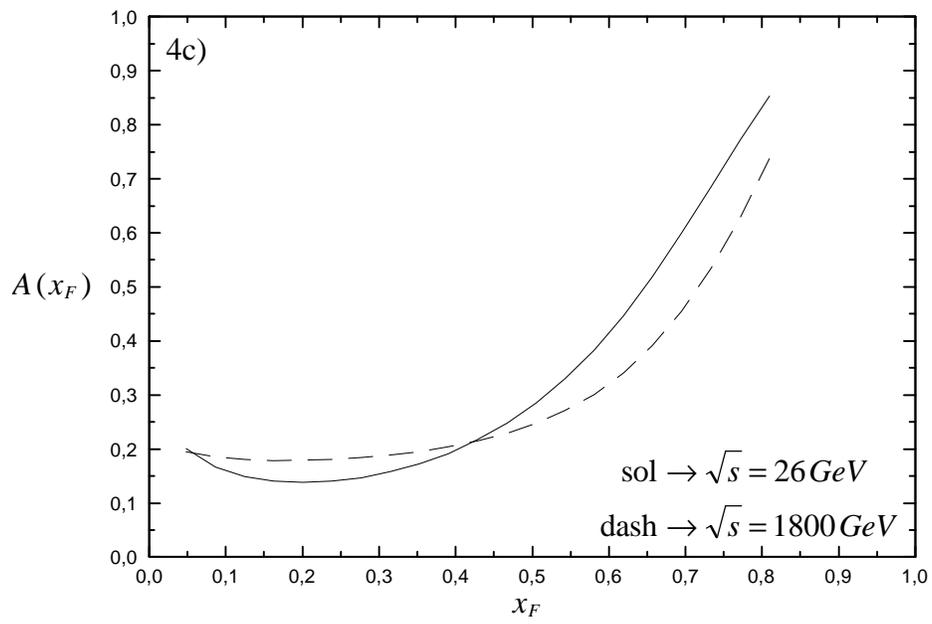
\item[{\bf Fig. 4}] $(a)$ Momentum distribution $\chi(x_{CF})$ of
the central fireball at $\sqrt{s}=26$ GeV (solid line) and at
$\sqrt{s}=1800$ GeV (dashed line) ; $(b)$ momentum distribution
$f_{LJ}(x_{L})$ of
the leading jet at $\sqrt{s}=26$ GeV (solid line) and at
$\sqrt{s}=1800$ GeV (dashed line); $(c)$ the asymmetry in pion-
proton collision calculated with the IGM: solid line corresponds
to $\sqrt{s}=26$ GeV with $\eta=0.7$ and dashed line corresponds
to $\sqrt{s}=1800$ GeV with $\eta=0.25$. In both cases $\xi=0.8$.
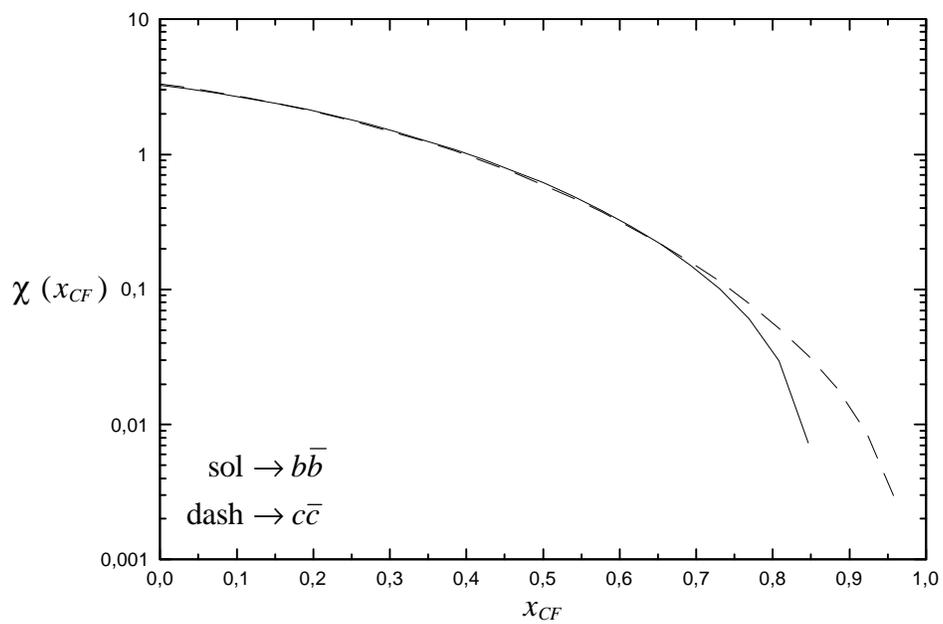
\item[{\bf Fig. 5}] $x_F$ distribution ($\chi(x_{CF})$) of centrally
produced $b-\overline b$ (solid line) and
$c-\overline c$ (dashed line) quark pairs.
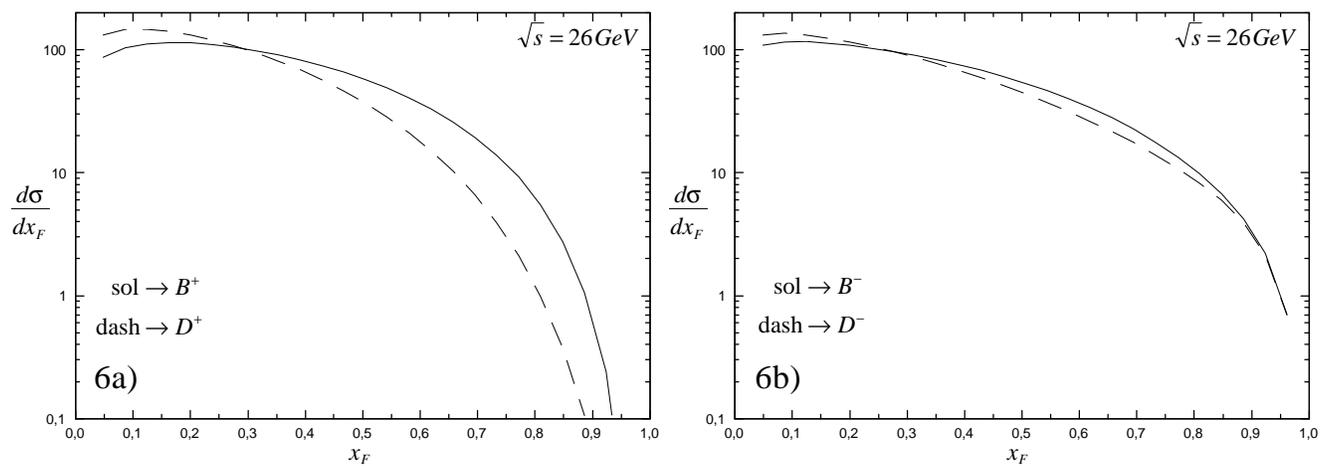
\item[{\bf Fig. 6}] $(a)$ $x_F$ distribution of nonleading $B$
(solid line) and $D$ (dashed line) mesons; $(b)$ the same as $(a)$
for leading $B$ (solid line) and $D$ (dashed line) mesons. The energy
is in both cases $\sqrt{s}=26$ GeV.
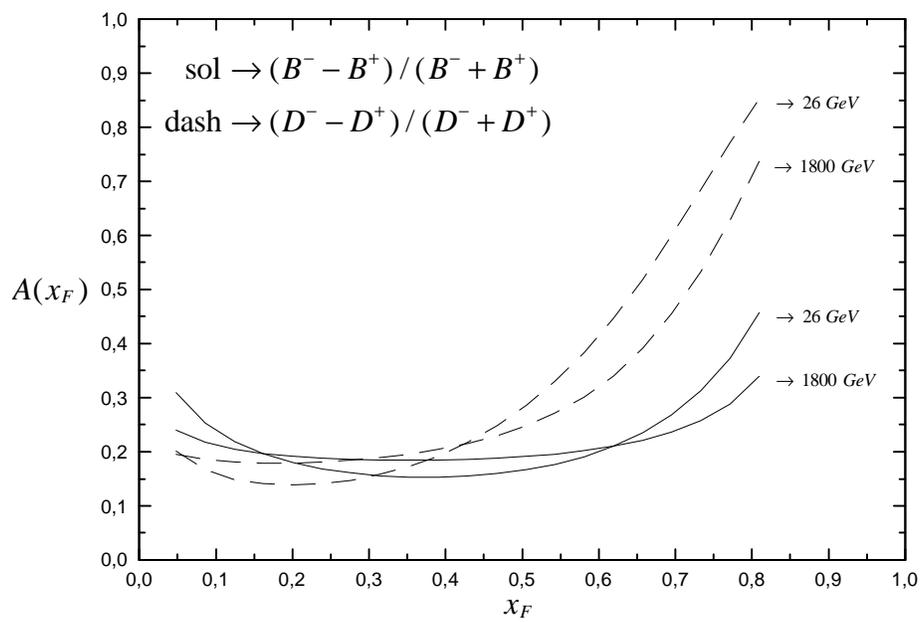
\item[{\bf Fig. 7}] $B^-$/$B^+$ (solid lines) and $D^-$/$D^+$
(dashed lines) asymmetries at $\sqrt{s}=$ 26  and 1800 GeV.

\end{itemize}


\begin{thebibliography}{99}
\bibitem{FOOT1} The energy dependence of the leading particle effect
                in charm production is relevant also for experimental
                reasons. The experiment E781, scheduled to run in
                the near future at Fermilab, is strongly based on the
                existence of this effect. Since it will use a pion
                beam of higher energies than presently available,
                predictions for the corresponding leading charm
                distributions are in order here.
\bibitem{FOOT2} Note that so far IGM has been applied only to
                reactions with nucleons. In order not to introduce
                new parameters when applying it to pion beams we are
                changing only the relevant cross sections entering
                it, namely $\sigma^{inel}_{NN}\rightarrow
                \sigma^{inel}_{\pi N}$, and gluon distributions.
\end{thebibliography}
\end{document}